\documentstyle[titlepage,12pt]{article}
\def\half{\frac{1}{2}}
\def\Rd{\dot{\cal{R}}}
\def\rd{\dot{r}}

\def\be{\begin{equation}}
\def\grad{\nabla}
\def\ee{\end{equation}}
\def\rq#1{(\ref{eq#1})}
\def\rsq#1{(\ref{seq#1})}
\def\SD{Schwarzschild-De Sitter }
\newcount\sectionnumber
\def\sect
{\global\equationnumber=0
\global\advance\sectionnumber by 1
\the\sectionnumber . }
\newcount\equationnumber
\def   \num
{\eqno{\global\advance\equationnumber by 1
\left(\the\sectionnumber .\the\equationnumber \right)}}
\begin{document}

\title{Matching Conditions and Gravitational Collapse
 in Two-Dimensional Gravity}

\author{R.B. Mann and S.F. Ross\\
        Department of Physics\\
        University of Waterloo\\
        Waterloo, Ontario\\
        N2L 3G1}
\date{May 27, 1992\\
WATPHYS TH-92/01}

\maketitle

\begin{abstract}
The general theory of matching conditions is developed for gravitational
theories in two spacetime dimensions. Models inspired from general
relativity and from string theory are considered. These conditions are used
to study collapsing dust solutions in spacetimes with non-zero
cosmological constant, demonstrating how two-dimensional black holes
can arise as the endpoint of such collapse processes.
\end{abstract}

\section{Introduction}

Although the study of matching conditions (phenomena at the boundary
between regions of space-time covered by different metrics) was first
considered around 1920, it has been a topic of considerable interest in
general relativity only since the influential paper by Israel in 1966
\cite{israel,chase,slake,Auril,llake}. The objective is to
describe the form of the surface separating the two regions, that is, to
give the spatial position of the surface as a function of its proper time.
In general relativity, this is accomplished by imposing matching conditions
on the metrics at the surface, where the metric on either side and the
stress-energy of the space-time are known \cite{israel}. This study has
several important applications in particle physics, astrophysics and
cosmology \cite{Auril}, as well as for gravitational collapse
\cite{llake,Weinberg}.

Interest in the study of such conditions in two spacetime dimensions
arises from recent work on theories of gravitation in this context
\cite{MST,Arnold,MSW,EDW,JurgJMP,L+R,RGRG}. Such theories have
a rich structure which can give rise to spacetimes with black holes,
thereby making them an interesting arena for the study of quantum
gravitational effects as their mathematical complexity is significantly
reduced from that of $(3+1)$ dimensional general relativity. The formation
of two-dimensional black hole spacetimes from collapsing distributions of
matter is of particular interest in this regard. Some \cite{symbh,CGHS}
(but not all \cite{Arnold}) of these static black hole spacetimes result
from the coupling of a dilaton field to the spacetime metric. The
presence of the dilaton necessitates a consideration of appropriate
matching conditions for the collapsing matter \cite{symbh}.

In this paper we formulate the $(1+1)$ dimensional analogue of the matching
conditions and apply them to both static spacetimes and to collapsing
matter in a variety of situations.  Specifically we consider theories of
gravitation in which a generic stress-energy tensor generates the spacetime
curvature, analogous to $(3+1)$-dimensional general relativity.  This may
be done via auxiliary fields \cite{semi}, leading to a theory in which the
Ricci scalar is set equal to the trace of the conserved stress-energy
tensor \cite{MST}, or by considering `string-inspired' models, in which
spacetime curvature arises due to a metric-dilaton coupling as well as due
to a generic stress-energy tensor \cite{MSW,symbh}. We show in each case
that the formation of black holes from collapsing dust proceeds in a
similar manner to the analogous collapse in general relativity.

\section{Two-Dimensional Gravity}

Before discussing the field equations, it is worthwhile to establish the
equivalent of the Gauss-Codazzi relations in two dimensions. As the
boundary $\Sigma$ between any two regions of $(1+1)$ dimensional spacetime
is in fact one-dimensional (a line), there is no intrinsic curvature, and
the extrinsic curvature may be given by the scalar
\be
K \equiv u^\alpha u^\beta \grad_\alpha n_\beta \label{eq003}
\ee
where $u^\alpha$ is the
tangent to $\Sigma$, and $n^\alpha$ is the normal. By applying
\be
R_{\alpha\beta\gamma\delta} = R g_{\alpha[\gamma}g_{\beta]\delta},
\label{eq004}
\ee
which is only true in two dimensions, and the fact that
$u^\alpha$ and $n^\alpha$ are orthogonal unit vectors, it is possible to
derive the equivalent of the Gauss-Codazzi relations,
\be
R = 2(n^\alpha\grad_\alpha K + \epsilon K^2),    \label{eq005}
\ee
where $\epsilon = -1$ if $\Sigma$ is timelike, $+1$ if it is spacelike. This
relates the only remaining degree of freedom in the curvature of the
spacetime, the Ricci scalar, to the extrinsic curvature of the surface.

Two-dimensional gravity must be founded on a different set of field
equations, as the result \rq{004} implies that Einstein's tensor vanishes
identically in two space-time dimensions,
\be
G_{\mu\nu} = R_{\mu\nu} - \half g_{\mu\nu} R \equiv 0. \label{eq24}
\ee
One proposed alternative is the theory
\be
R-\Lambda = 8\pi GT,    \label{eq25}
\ee
which is the simplest non-trivial analogue of general relativity that may
be constructed in two dimensions \cite{Rsimp} .  This equation
derived from the action \cite{semi}
\be
S=\int d^2x\sqrt{-g}g^{\mu\nu}
\left(\half\partial_\mu\Psi\partial_\nu\Psi+\Psi
R_{\mu\nu}+\half g_{\mu\nu}\Lambda \right) + {\cal L}_M,          \label{eq26}
\ee
which also implies conservation of energy,
\be
T^{\mu\nu}_{\ \ ;\nu} = 0,  \label{eq27}
\ee
if we define $\delta{\cal L}_M\equiv -8\pi G\sqrt{-g}T_{\mu\nu}\delta
g^{\mu\nu}$. For reasonable two-dimensional stress-energy tensors, the
field equation \rq{25} qualitatively reproduces many of the solutions of
Einstein's equations, including gravitational collapse, a cosmological
solution, a post-Newtonian expansion, and black holes \cite{Arnold,MST},
plus a number of other similarities in the semiclassical regime
\cite{MST,semi,SharTomRobb}.  We shall refer to this as the ``R=T'' theory.

The study of matching conditions in this theory leads to some results
similar to general relativity, although we do see differences, particularly
in the physical interpretation. The analogue of the Lanczos equations for
this field equation is \cite{who}
\be
4\pi G S = [K],   \label{eq006}
\ee
and integration of the field equation \rq{25} using \rq{005} gives
\be
S = \lim_{\varepsilon\rightarrow 0} \int_{0}^{\varepsilon} T dn,  \label{eq007}
\ee
that is, $S$ is the integral of the trace of the stress-energy through the
surface which is analogous to the relation
\be
S_{ij}= \lim_{\varepsilon\rightarrow 0} \int_{0}^{\varepsilon}
T_{ij} dn,\label{eq9}
\ee
in general relativity \cite{israel}.

Another two-dimensional theory of recent interest has been developed from a
non-critical string theory in two target-space dimensions \cite{MSW,EDW},
although the most interesting solution, a metric having the form of an
asymmetric black hole,
\be
ds^2 = -(1-ae^{-Qx})dt^2 + \frac{dx^2}{1-ae^{-Qx}}      \label{eq28}
\ee
was first found as the solution to a scale-invariant higher-derivative theory
of gravity \cite{JurgJMP}, and has also been found to be a solution to $c=1$
Liouville gravity in two dimensions \cite{L+R}. The effective target space
action for the string theory is \cite{MSW}
\be
S = \int d^2x e^{-2\Phi} \sqrt{-g}(R-4(\grad\Phi)^2+c). \label{eq29}
\ee
{}From a gravitational point of view, the asymmetry of the solution \rq{28}
about the origin is somewhat objectionable, as it is difficult to understand
how such a solution could arise from gravitational collapse of a
distribution of matter confined to some finite spatial region (for an
alternative viewpoint see \cite{CGHS}). It is possible \cite{symbh} to
derive a symmetric solution locally identical to \rq{28} by assuming the
presence of an appropriate source of stress-energy centered about the
origin. This may be done by introducing a corresponding matter term ${\cal
L}_M$ in the action \rq{29} which may be thought of as modelling some
unknown higher-order effects. It may also be shown that the action with
matter term is equivalent to that of a massive scalar field $\psi = e^{-
\Phi}$ non-minimally coupled to curvature \cite{symbh}, allowing a less
ambiguous interpretation of the matter term.

We therefore propose the revised action
\be
S = \int d^2x \sqrt{-g}\left\{ e^{-2\Phi}(R-4(\grad\Phi)^2+c)
+ {\cal L}_M \right\} .
\label{eq30}
\ee
This leads to the field equations
\be
R_{\mu\nu} -2\grad_\mu\grad_\nu \Phi
= 8\pi Ge^{2\Phi} T_{\mu\nu}, \label{eq31}
\ee
and
\be
R-4(\grad\Phi)^2+4\grad^2\Phi+c = 0.     \label{eq32}
\ee
This system is also quite rich in structure, having
a cosmological solution, a post-Newtonian expansion, the
symmetric black hole solution
\be
ds^2 = -(1-ae^{-Q|x|})dt^2 + \frac{dx^2}{1-ae^{-Q|x|}}      \label{eq33}
\ee
and gravitational collapse. These have been studied in detail in
\cite{symbh,postnewt}.
Although these field equations sacrifice some of the simplicity of the field
equation \rq{25}, they remain substantially simpler in practical terms than
general relativity. The analogues of the Lanczos equations for these field
equations are
\be
8\pi GS_{00} = \epsilon [K],     \label{eq34}
\ee
\be
4\pi GS_{01} = [u^\alpha\grad_\alpha \Phi],      \label{eq35}
\ee
\be
8\pi GS_{11} = -\epsilon[K] + 2[n^\alpha\grad_\alpha \Phi],      \label{eq36}
\ee
in natural coordinates \cite{israel} where $x^0$ and $x^1$ are the geodesic
distances along $u^\alpha$ and $n^\alpha$ respectively. These equations
introduce the possibility of non-zero elements of $S_{ij}$ in the normal
direction,  in contrast to the situation in general relativity.

Integration of the field equations (\ref{eq31},\ref{eq32}) using
the relation between the Ricci scalar and extrinsic curvature \rq{005} gives
\be
S_{ij} = \lim_{\varepsilon\rightarrow 0}
\int_{0}^{\varepsilon} e^{2\Phi} T_{ij} dn,   \label{eq37}
\ee
that is, the surface stress-energy is the integral of the stress-energy
multiplied by a weighting factor through the surface in natural
coordinates which is the analogue of \rq9 in this case.

\section{Matching: The Static Case}

In the R=T theory, the metric
\be
ds^2 = -f(|x|) dt^2 + f(|x|)^{-1} dx^2  \label{eq38}
\ee
where
\be
f(|x|) = -\frac{\Lambda}{2}x^2 + 2M|x| -C \label{eq39}
\ee
is a solution of the field equation \rq{25} for $8 \pi GT = 4M\delta(x)$,
that is, it is the general vacuum solution with a point source of matter
and a cosmological constant. It is easy to see that this metric describes
black hole solutions and cosmological event horizons, as the coordinate
singularities of the metric are given by the condition $f(|x|)=0$, giving
pairs of horizons on either side of the point singularity. We therefore say
that this metric represents the analogue of the \SD solution in this  two
dimensional theory.  Note however that the Cosmic Censorship principle does
not apply, as, for given $\Lambda$ and $M$, there is some $C$ such that
$f(|x|)=0$ has no positive roots, and thus there can be uncloaked
singularities.

As any two-dimensional space-time metric has only one degree of freedom,
any symmetric static metric can be written in the form \rq{38}. It is
therefore very useful to consider the problem analogous to the boundary
between two \SD metrics \cite{Auril},
\be
ds^2_\pm = -f_\pm dt^2_\pm +\frac{dx^2}{f_\pm}, \label{eq40}
\ee
where the physical example we consider is the boundary between two
solutions of the form
\be
f_\pm = -\half\Lambda_\pm x^2 + 2M_\pm |x| -C.   \label{eq41}
\ee
The metric intrinsic to $\Sigma$ is
\be
ds^2_{\Sigma} = -d\tau^2  \label{eq008}
\ee
in all two-dimensional systems. Matching the metrics at the surface $\Sigma$
described by $|x| = {\cal R}(\tau)$ gives
\be
f_\pm^{-1}\Rd^2-f_\pm\dot{t}_\pm^2 = -1,       \label{eq42}
\ee
where the overdot denotes the derivative with respect to proper time, as
before. Assuming the surface is timelike, the tangent vector
will be a timelike unit vector,
\be
u^\mu u^\nu g_{\mu\nu} = -1     \label{eq43}
\ee
implying by \rq{42} that
\be
u^\mu = (\dot{t}_\pm, \Rd).     \label{eq009}
\ee
Similarly, by using the orthogonality of the tangent and normal vectors,
$u^\mu n_\mu = 0$, and the fact that $n_\mu$ is also a unit vector,
\be
n_\mu n_\nu g^{\mu\nu} =1, \label{eq44}
\ee
we find
\be
n_\mu= (-\Rd,\dot{t}_\pm).   \label{eq010}
\ee
Applying the condition \rq{42} and the two forms (\ref{eq009},\ref{eq010})
to the definition \rq{003} of the extrinsic curvature yields
\be
K_\pm = -\frac{\ddot{{\cal R}} +\half f'_\pm}{\sqrt{\Rd^2+f_\pm}},
\label{eq45}
\ee
which may be simplified to
\be
K_\pm  = -\frac{d}{d{\cal R}} \sqrt{\Rd^2 + f_\pm}.     \label{eq46}
\ee
Substituting this form of $K$ in the Lanczos equation $[K] = 8 \pi GS$ gives
\be
\dot{F}_+ - \dot{F}_- = -8 \pi GS \Rd, \label{eq47}
\ee
$F_\pm=\sqrt{\Rd^2+f_\pm}$ as before, analogous to the situation in general
relativity \cite{chase}.  However, we do not have the additional
information (which comes from the angular components in higher dimensions)
that enabled us to obtain the first integral previously.
If we define $\sigma$ by
\be
\frac{d(\sigma {\cal R})}{d\tau} = 2S \Rd     \label{eq48}
\ee
then we obtain
\be
F_+ - F_- = -4\pi G\sigma {\cal R} - c,  \label{eq49}
\ee
with ${c}$ a constant of integration. This equation is identical in
form to that which occurs in general relativity \cite{chase}. However the
physical interpretation of $\sigma$ is unclear in this case, and (in
contrast to general relativity \cite{chase}) we cannot eliminate the
constant of integration ${c}$.  This constant means two initial
conditions are required to give a unique solution, where only one was
necessary in general relativity. In the special case of the thin shell of
dust in a vacuum, $S$ is a constant, the surface density, and we get
$\sigma=2S$ from the definition \rq{48}. In general relativity, $\sigma$ is
defined to be the surface density, so this result is encouraging.

Turning now to the string-motivated theory described by (\ref{eq30}),
if we consider the case of the symmetric static metric
\be
ds^2_\pm = -f_\pm dt^2_\pm +\frac{dx^2}{f_\pm} \quad , \label{eq50}
\ee
the
development could be followed through in much the same way as it was above
for the R=T theory. However, in the case of (\ref{eq31},\ref{eq32}), we
have another important matching condition. We must require that the dilaton
field $\Phi$ be continuous, as otherwise we would need an infinite surface
density of the associated charge on the surface (this is similar to the
requirement that the scalar potential $V$ be continuous in classical
electromagnetism).  We have two important static metrics in this theory,
which are (written in the form \rq{50})
\be
f_+ = 1-ae^{-Q_+|x|},  \label{eq51}
\ee
the vacuum solution developed in \cite{symbh}, for which we find
\be
\Phi_+ = -\frac{Q_+|x|}{2},        \label{eq52}
\ee
and
\be
f_- = 1- \frac{k}{1+\cosh Q_-x}  \label{eq53}
\ee
which is actually a proposed vacuum solution of the string theory with a
non-zero tachyon field \cite{VerDijk}, with the associated dilaton field
\be
\Phi_- =
\ln\left[Q_-\sqrt{2/k}\sinh(\frac{Q_-}{2}|x|)\sqrt{\sinh^2(\frac{Q_-}{2}|x|)
+1}\right].    \label{eq54}
\ee
Thus, if we regard these two solutions as the outside and inside metrics for
some surface layer, requiring continuity of the dilaton field
$\Phi_+({\cal R})=\Phi_-({\cal R})$ gives the equation
\be
e^{-Q_+R} =\frac{2Q_-^2}{k}\sinh^2(\frac{Q_-}{2}{\cal R})
\left(\sinh^2(\frac{Q_-}{2}{\cal R})+1\right),  \label{extra}
\ee
which has only the trivial solution ${\cal R}(\tau) = C$,  where $C$ is a
constant determined in terms of the parameters on both sides. It may be the
case that there is no solution, as (\ref{extra}) may not have a positive
real root. The constant solution does not imply that there is
no stress-energy on the surface; instead
\be
K_\pm = -\frac{d\sqrt{f_\pm}}{d{\cal R}},    \label{eq55}
\ee
as $\Rd=0$, and thus the first ``Lanczos'' equation \rq{34} implies
\be
8\pi GS_{00} = -\frac{d}{d{\cal R}}(\sqrt{f_+} -\sqrt{f_-}).   \label{eq56}
\ee
We may make the obvious generalization that for any matching problem in
which the dilaton field is static on both sides of the surface, requiring
continuity of the dilaton field will give at most the trivial solution
${\cal R}(\tau)=C$, and some constant stress-energy on the surface.  A
complete solution would require matching of the tachyon field as well, but
we shall not consider this problem any further.

\section{Matching: The Collapsing Case}

Consider a non-static metric written in comoving coordinates
\be
ds^2 = -dt^2+a^2(t)dy^2,         \label{eq57}
\ee
where the cosmic scale factor $a(t)$ is an arbitrary function representing
the one degree of freedom of the metric. We take the position of the
surface $\Sigma$ to be $|y|=r(\tau)$ in this metric. This metric represents
the cosmological solution, as the length scale will vary in time; in
two dimensions open and closed universes do not yield distinct metrics
\cite{Arnold}. In the two-dimensional theories described
above, the exact solutions to date can be written either in this form or in
the static form \rq{38}; in particular, the interior metric of collapsing
dust is written in this form with the spatial coordinate taken to be
co-moving, implying $\rd=0$.

If the metric on one side (say $V_-$) of a surface has the form \rq{57},
matching it to the intrinsic metric of the surface gives the condition
\be
-\dot{t}_-^2 + a^2\rd^2 =-1.     \label{eq58}
\ee
Assuming the surface is timelike, the expressions \rq{009} and \rq{010} for
$u^\alpha$ and $n_\alpha$ are valid, and we can explicitly evaluate the
extrinsic curvature, giving
\be
K_- = -\frac{\ddot{r}}{\dot{t}_-}-2\frac{a'}{a}\rd.      \label{eq59}
\ee
As this does not appear to be the derivative of a function, that is $K_- \neq
df/dr$ for any $f$, we cannot integrate the Lanczos equation as we did
previously. Therefore, this is as far as we can carry the general argument,
and further discussion requires that we specify $a(t)$.

We therefore turn to the special case of a collapsing ``ball'' of dust in one
spatial dimension. The study of a ball of collapsing dust in general
relativity provides an idealized model of the physical collapse of a star
to form a black hole. Here we consider the analogue of this problem
in two-spacetime dimensions to see whether or not the 2D black holes
studied previously \cite{MST,MSW,RGRG} can arise as the endpoint of such
a collapse.

Although a ball of collapsing dust in a spacetime with vanishing
cosmological constant  was studied for these reasons early in the
development of the R=T theory \cite{Arnold}, it has been recently suggested
\cite{kriele} that as the space-time outside this dust is flat, it
represents a very special case, and is not a reliable model for the
corresponding interactions in general relativity. We therefore consider
collapsing dust in a spacetime with non-zero cosmological constant. As we
shall see, the qualitative properties of the collapse are the same as in
ref. \cite{Arnold}, and all the results reduce to the previous ones when
$\Lambda \rightarrow 0$.

The dust is falling freely under the
influence of purely gravitational forces, so we may make it the basis of a
comoving coordinate system. We then have a Robertson-Walker metric on the
interior region,
\be
ds^2 = -d\tau^2 + a^2(\tau) dy^2,    \label{seq3}
\ee
where $y$ is the comoving spatial coordinate, $\tau$ is the proper time of
the dust, and $a$ is the scale factor.
The trace of the stress-energy in the interior region is $T=-\rho$, where
$\rho(\tau)$ is the density of the dust. The conservation law \rq{27} thus
implies that $\rho a = \rho_0 a_0$, where $\rho_0$ is the initial density
of the dust, and $a_0$ is the initial scale factor \cite{Arnold}. The field
equation then becomes
\be
\ddot{a} = -4 \pi G \rho_0 a_0 + \frac{\Lambda}{2} a, \label{seq5}
\ee
where the overdot denotes $d/d\tau$, and we impose the initial conditions
\be
a_0=a(0) =1, \dot{a}(0)=0,   \label{seq6}
\ee
that is, we take the dust to be initially at rest with $a$ at its maximum,
and scale $a$ to its initial value. The solution for $a$ with these initial
conditions can be easily seen to be
\be
a(\tau)=(1-\frac{8\pi G\rho_0}{\Lambda})\cosh(\sqrt{\frac{|\Lambda|}{2}}\tau)+
\frac{8\pi G\rho_0}{\Lambda} \label{seq7}
\ee
if $\Lambda$ is positive, and
\be
a(\tau)=(1-\frac{8\pi G\rho_0}{\Lambda})\cos(\sqrt{\frac{|\Lambda|}{2}}\tau)+
\frac{8\pi G\rho_0}{\Lambda} \label{seq8}
\ee
if $\Lambda$ is negative. As $\Lambda \rightarrow 0$
\be
a(\tau) \rightarrow 1 - 2\pi G\rho_0 \tau^2,  \label{seq9}
\ee
which is the solution found in \cite{Arnold}.

To make the form of
the solution more obvious, we write the interior metric as
\be
ds^2 = -d\tau^2 + ((1-b)\cosh(\sqrt{\frac{\Lambda'}{2}}\tau)+b)^2 dy^2
\label{seq10}
\ee
if $\Lambda$ is positive, and
\be
ds^2 = -d\tau^2 +((1+b)\cos(\sqrt{\frac{\Lambda'}{2}}\tau)-b)^2 dy^2
\label{seq11}
\ee
if $\Lambda$ is negative, where we have defined $\Lambda' =|\Lambda|$, and
$b = 8\pi G\rho_0/\Lambda'$.  For a collapsing dust, we must have
$a(\tau)=0$ at some finite time $\tau$. For $\Lambda$ negative,
\be
\tau= \sqrt{\frac{2}{\Lambda'}} \cos^{-1}\left(\frac{b}{b+1}\right),
\label{sequ11}
\ee
but when $\Lambda$ is positive, we must have
\be
b>1 \Rightarrow \rho_0>\frac{\Lambda}{8 \pi G} \label{sequ1}
\ee
for \rsq{10} to correspond to a collapsing dust. If this condition is not
satisfied, the gravitational collapse of the dust will be overcome by the
general expansion of the space-time caused by the cosmological constant.
When (\ref{sequ1}) is satisfied, $a(\tau)=0$ at
\be \tau= \sqrt{\frac{2}{\Lambda'}} \cosh^{-1}\left(\frac{b}{b-
1}\right).\label{sequ12}
\ee
Note that $\tau \rightarrow (2\pi G\rho_0 )^{-1/2}$ as $\Lambda \rightarrow
0$ in both cases, reproducing the previous result \cite{Arnold}.

In the exterior coordinates, where the stress-energy trace $T=0$, the
metric is the two-dimensional \SD analogue \rq{38}. In the matching, we
treat the mass $M$ and constant $C$ as adjustable parameters.

We consider next the problem of extending the outside co-ordinates to the
inside of the dust. In the integrating factor
technique, we attempt to construct a continuous transformation from the
interior to the exterior coordinates over the whole of the interior dust-
filled region \cite{Weinberg}. That is to say, we attempt to write the
interior metric in the form
\be
ds^2 = -B(t,x)dt^2 + A(t,x){dx^2}
\label{seq14}
\ee
where $x$ and $t$ are the same as before, and $B$ and $A$
must satisfy the boundary conditions
\be
B(t,{\cal R}) = f({\cal R}), \quad  A(t,{\cal R})  =
\frac{1}{f({\cal R})} \label{sequ13}
\ee
at the dust edge, whose position is
$|y|=r$ in the interior coordinates and $|x|={\cal R}$ in the exterior
coordinates. We assume that
\be
dt= \eta (\dot{S} d\tau+ S' dy), x = a(\tau) y    \label{seq15}
\ee
where $\eta$ is an integrating factor.
Requiring that the forms \rsq{14} and \rsq3 of the interior metric be
equivalent then gives
\be
B\eta^2\dot{S} S' = Aya\dot{a}, \label{seq16}
\ee
\be B\eta^2\dot{S}^2 - Ay^2\dot{a}^2 =1, \label{seq17} \ee \be B\eta^2S'^2-
Aa^2 =-a^2,  \label{seq18}
\ee
and
\rsq{17} and \rsq{18} imply
\be A= \frac{1}{1-y^2\dot{a}^2}.     \label{seq18a}
\ee

If the cosmological constant vanishes, it is possible to solve for $S$ and
$\eta$, and explicit forms of $A$ and $B$ are given in \cite{Arnold}. In
the more general case, we were able to find $S$ by assuming
$S=f(y)g(a(\tau))$, but we were unable to find a suitable expression for
$\eta$. As a transformation of this kind was found in the case $\Lambda=0$,
we expect that it is still in principle possible here, since there should
be no significant change in the behavior of the dust edge. We therefore
assume
\be
f({\cal R}) = \frac{1}{A} = 1-r^2\dot{a}^2,    \label{seq19}
\ee
which is necessary for the
transformation to be continuous at the dust edge,  and ${\cal R}=ra(\tau)$,
that
is, the position of the dust edge in the exterior coordinates is the proper
distance from the origin to the dust edge. Note that with the latter
assumption, we can interpret the boundary conditions \rsq6 as representing
a ball of dust with initial radius ${\cal R}_0=r$ initially at rest in the
exterior coordinates, and we see that $a=0$ corresponds to the collapse of
the dust to ${\cal R}=0$.

We next show that the above conditions imply that the dust edge is in fact a
boundary surface in two dimensions. The extrinsic curvature in the interior
region is given by \rq{59}, and as the interior coordinates are co-moving,
$\rd=0$, and thus $K_-=0$. The extrinsic curvature in the exterior coordinates
is given by \rq{46},
\be
K_+  = -\frac{d}{d{\cal R}} \sqrt{\Rd^2 + f}.     \label{eq46again}
\ee
If we now set ${\cal R}=ra(\tau)$, $f({\cal R}) = 1-r^2\dot{a}^2$, we see
that $\Rd^2+f({\cal R})=1$, and thus $K_+=0$. Thus, if these conditions are
satisfied, the dust edge will be a boundary surface. This is important, as
the dust edge in general relativity is a boundary surface, and thus a
collapse that required a surface shell of matter would not be a good model
for the collapsing dust in four dimensions. Finally, we should note that
\rsq{19} is not a necessary condition for the dust edge to be a boundary
surface, as we only need $\Rd^2+f({\cal R})=c$, $c$ some arbitrary constant, to
obtain $K_+=0$.

Applying the conditions \rsq{19} and ${\cal R}=ra(t)$ to the interior
metrics \rsq{10} and \rsq{11}, and the exterior metric \rq{38}, we find
\be
M = 2 \rho_0 r, \quad
C = 1-r^2(8\pi G\rho_0 - \frac{\Lambda}{2}). \label{seq26}
\ee
This mass identification is hardly surprising, as it simply means that a
line of dust of density $\rho_0$ and length $2r$ is equivalent to a point
source of mass $2\rho_0 r$. The identification of the constant of
integration is relatively unimportant, as it has no physical significance.
Note that only the mass identification is required for the dust edge to be
a boundary surface, that is, it is equivalent to the necessary condition
$\Rd^2+f({\cal R})=c$.

As positive and negative cosmological constants give slightly different
results we treat them separately. For positive $\Lambda$, the matching
conditions are satisfied if \rsq{10} is matched to \rq{38} with
\be
f(x) = \Lambda' br |x| - \frac{\Lambda'}{2}x^2+1
-\frac{\Lambda'}{2}r^2(2b-1). \label{seq27}
\ee

There is no Birkhoff theorem in two dimensions \cite{SharTomRobb}, so we
must ask under what conditions the collapse of the dust leads to a black
hole. The positions of the event horizons are given by the roots of
\rsq{27}, as these are where the signature of the metric \rq{38} changes
sign, so
\be
|x_h| = br \pm \sqrt{(b-1)^2r^2+\frac{2}{\Lambda'}}.   \label{seq28}
\ee
There is always at least one root, but we interpret this larger root as the
cosmological event horizon due to the expansion of space-time caused by the
positive cosmological constant. Thus, an event horizon will form around the
collapsing dust only if the smaller root is positive, that is, if
\be
b>\frac{1}{\Lambda' r^2}+\half \Rightarrow \rho_0 > \frac{1}{8\pi Gr^2} +
\frac{\Lambda'}{16\pi G}. \label{seq29}
\ee
This condition reduces to the one in \cite{Arnold} if $\Lambda=0$.
Note that it is possible to satisfy this condition on the initial
density, but not the condition (\ref{sequ1}). This, however, is physically
meaningless, as in this case the dust would not collapse, and the Schwarzschild
radius given by \rsq{28} would never lie within the range of the exterior
coordinates. Alternatively, if (\ref{sequ1}) is satisfied, but \rsq{29} is
not, collapse occurs to a naked point source. Both (\ref{sequ1}) and
\rsq{29} must be satisfied for a black hole to form.

If an event horizon does form, we would now like to know when it
forms, that is, we would like to find the comoving time $\tau_h$ at which the
event horizon and the dust edge coincide. This is found by substituting
$x_h = ra(\tau_h)$ in \rsq{28}, which gives
\be
\tau_h= \sqrt{\frac{2}{\Lambda'}}\sinh^{-1}\left(\sqrt{\frac{2}{\Lambda'}}
\frac{1}{r(b-1)}\right). \label{seq30}
\ee
Note that the comoving time at which the horizon forms is finite, and
$\tau_h \rightarrow 1/4\pi G\rho_0 r$ as $\Lambda \rightarrow 0$, so we recover
the previous result \cite{Arnold}.

A light signal emitted from the surface at time $t$ obeys the null condition
\be
\frac{dx}{dt} = \Lambda' br|x| - \frac{\Lambda'}{2}x^2+1-\frac{\Lambda'}{2}
r^2(2b-1) \label{seq31}
\ee
and arrives at a point $\tilde{x}$ at time
\begin{eqnarray}
\tilde{t} & = & t + \int_{ra(\tau)}^{\tilde{x}} \frac{dt}{dx} dx,\nonumber \\
 & = & t + \frac{2}{\Lambda'\sqrt{(b-1)^2r^2+2/\Lambda'}}\tanh^{-1}\left[
 \frac{|x|-br}{\sqrt{(b-1)^2r^2+2/\Lambda'}}\right]_{ra(\tau)}^{\tilde{x}}
 \!\!\!\!\!\! ,\label{seq32}
\end{eqnarray}
and thus $\tilde{t} \rightarrow \infty$ as $ra(\tau) \rightarrow x_h$ given by
\rsq{28}, that is, as $\tau \rightarrow \tau_h$,
so the collapse to the Schwarzschild radius appears to take an infinite amount
of time, and the collapse to ${\cal R}=0$ is unobservable from outside, as in
\cite{Arnold}. The form of $\tilde{t}$ also reduces to the previous one as
$\Lambda \rightarrow 0$.

The comoving time interval $d\tau$ between emissions of wave crests
is equal to the natural wavelength $\lambda$ that would be emitted in the
absence of gravitation, and the interval $d\tilde{t}$ between arrivals of wave
crests is the observed wavelength $\tilde{\lambda}$. Thus the red shift of
light
from the dust edge is
\be
z = \frac{d\tilde{t}}{d\tau}-1 = \frac{1}{1+r\dot{a}(\tau)} -1,
\label{seq33}
\ee
and
\be
r\dot{a}(\tau_h) = r(1-b)\sqrt{\frac{\Lambda'}{2}}\sinh\left(
\sqrt{\frac{\Lambda'}{2}}\tau_h\right) = -1,   \label{sequ14}
\ee
therefore $z \rightarrow \infty$ as $\tau \rightarrow \tau_h$, as before
\cite{Arnold}. Thus, the collapsing fluid will fade from sight, as the red
shift
of light from its surface diverges.

For negative $\Lambda$, the matching conditions are satisfied if \rsq{11} is
matched to \rq{38} with
\be
f(x) = \Lambda' br|x|+ \frac{\Lambda'}{2}x^2+1-\frac{\Lambda'}{2}r^2(2b+1).
\label{seq34}
\ee
The position of the event horizon is now given by the root of \rsq{34},
which gives
\be
|x_h| = -br \pm \sqrt{(b+1)^2r^2-\frac{2}{\Lambda'}},  \label{seq35}
\ee
and we see that there is at most one positive root of \rsq{34}. This is not
surprising, as with $\Lambda$ negative, the space-time is contracting, and
there is no cosmological horizon. This Schwarzschild radius is positive if
\be
b>\frac{1}{\Lambda' r^2}-\half \Rightarrow \rho_0>\frac{1}{8\pi Gr^2}-
\frac{\Lambda'}{16\pi G}. \label{seq36}
\ee
This also reduces to the condition in \cite{Arnold} if $\Lambda=0$.

The comoving time at which the horizon forms may be found by substituting
$x_h= ra(\tau_h)$ in \rsq{35}, which gives
\be
\tau_h = \sqrt{\frac{2}{\Lambda'}}\sin^{-1}\left(\sqrt{\frac{2}{\Lambda'}}
\frac{1}{r(b+1)}\right).   \label{seq37}
\ee
As before, $\tau_h$ is finite, with limit
$\tau_h \rightarrow 1/4\pi G\rho_0 r$ as
$\Lambda \rightarrow 0$, so we recover the previous result \cite{Arnold}.
The null condition for a light signal emitted from the dust edge is
\be
\frac{dx}{dt}=\Lambda'
br|x|+\frac{\Lambda'}{2}x^2+1-\frac{\Lambda'}{2}r^2(2b+1)
\label{seq38}
\ee
so light emitted at $(x,t)$ arrives at $\tilde{x}$ at time
\begin{eqnarray}
\tilde{t}& = &t + \int_{ra(\tau)}^{\tilde{x}} \frac{dt}{dx}dx \nonumber \\
  & = &t -\frac{2}{\Lambda'\sqrt{(b+1)^2r^2-2/\Lambda'}}\tanh^{-1}\left[
  \frac{|x|+br}{\sqrt{(b+1)^2r^2-2/\Lambda'}}\right]_{ra(\tau)}^{\tilde{x}}
  \!\!\!\!\!\! ,\label{seq39}
\end{eqnarray}
and thus $\tilde{t} \rightarrow \infty$ as $ra(\tau) \rightarrow x_h$ given by
\rsq{35}, that is, as $\tau \rightarrow \tau_h$. The collapse to
the Schwarzschild radius appears to take an infinite amount of time, as before.
The form of $\tilde{t}$ also reduces to the previous one as $\Lambda
\rightarrow
0$.

The red shift of light from the horizon is still given by \rsq{33}, and we have
\be
r\dot{a}(\tau_h)=
-r(1+b)\sqrt{\frac{2}{\Lambda'}}\sin\left(\sqrt{\frac{2}{\Lambda'}}
\tau_h\right) = -1, \label{sequ15}
\ee
so $z \rightarrow \infty$ as $\tau \rightarrow \tau_h$, the red shift of
light from the surface diverges as the horizon forms. We should note
finally that \rsq{33} reduces to the expression in \cite{Arnold} for the
red shift when the cosmological constant vanishes.

The case of collapsing dust in the string-motivated theory was studied in
\cite{symbh}. We include the results here in the interests of completeness,
and because they provide another example of the effects of requiring
continuity of the dilaton field. The interior metric for the dust is of the
form
\be
ds^2_- = -d\tau^2+a^2(\tau)dy^2,   \label{eq63}
\ee
where the $y$ coordinate is comoving with the dust. The interior dilaton field
is
\be
\Phi_-(\tau) = \Phi_0 - \ln\left(\cos(\frac{Q}{2}\tau)\right).
\label{eq64}
\ee
The exterior metric is the vacuum solution \rq{33},
\be
ds^2_+ = -(1-ae^{-Q|x|})dt^2 + \frac{dx^2}{1-ae^{-Q|x|}},      \label{eq65}
\ee
and the exterior dilaton field is
\be
\Phi_+  = -\frac{Q}{2}x.   \label{eq66}
\ee
We require continuity of the dilaton field through the surface, which gives the
equation
\be
{\cal R}(\tau)
= x_0 + \frac{2}{Q}\ln\left(\cos(\frac{Q}{2}\tau)\right) \label{eq67}
\ee
in the exterior coordinates. There will in general be a non-zero
stress-energy  on the surface of the dust in the string theory,
\be
8\pi GS_{ij} = \left(\begin{array}{cc} -[K] & 0 \\ 0 & -[K] \end{array}\right),
\label{eq68}
\ee
and there is a surface dilaton charge, a feature that differs from both the
R=T case and that of general relativity.

\section{Conclusions}

We have shown that the endpoint of dust collapse in both the R=T and the
string-motivated two dimensional theories of gravity considered here
yield the static black
holes studied previously \cite{MST,MSW}. The general results
qualitatively parallel the four-dimensional case, although the technical
details are quite different. In particular, we have \rq{005}, which has the
same content as the Gauss-Codazzi equations. The ``Lanczos'' equations in
the two cases take almost the same form as in general relativity, and give
the same result that $S_{ij}$, referred to as a surface stress-energy, is
the integral of the volume stress-energy through the surface in natural
coordinates, although this is multiplied by an exponential of the
dilaton field  in the string-motivated
case. It is worth remarking that in the latter theory, the normal
components of $S_{ij}$ are non-zero, unlike general relativity and the
linear theory.

At the boundary between two static metrics in the R=T theory, we find that
the equation for ${\cal R}(\tau)$ is the same as it was in general relativity,
with the problem being reduced to solving the differential equation
\rq{49}. However, the interpretation of this result is quite different:
$\sigma$ does not have a simple physical interpretation, and there is an
additional arbitrary constant in the solution. On the other hand the string
motivated system of equations (\ref{eq31},\ref{eq32}) has a
physically significant auxiliary field, the dilaton field $\Phi$, and we
find that requiring that $\Phi$ be continuous gives us the function
$R(\tau)$ describing the surface in both cases of interest. As the dilaton
field represents a physical potential in the string theory, we must require
that it be continuous if we wish to have only finite effects on the
surface.

For the R=T theory we have extended the collapsing dust scenario of ref.
\cite{Arnold} to include spacetimes with a non-zero cosmological constant.
The solution with $\Lambda$ positive will collapse only if the initial
density $\rho_0$ exceeds a critical value given by (\ref{sequ1}), since the
gravitational attraction of the dust must overcome the cosmological
expansion of the space-time. The condition \rsq{19} is sufficient to make
the dust edge a boundary surface, although it is not necessary. It is
equivalent to the mass and constant identifications \rsq{26}. This is the
mass we would expect for the dust, and it is equivalent to the necessary
condition for the dust edge to be a boundary surface. Horizons
will form as the dust collapses only if the conditions \rsq{29} and
\rsq{36} on the initial density of the dust are satisfied.  We also find
that when the horizon forms, it does so in finite comoving time and
infinite coordinate time, and that the red shift of light from the dust
edge diverges as the horizon forms.

As the qualitative results of the study
of the collapsing dust are the same as those of the previous study
\cite{Arnold}, we conclude that black holes may be successfully modelled in
two dimensions. Of greater interest, of course, is the inclusion of quantum
effects in these processes. The simplicity of the theories described here
make them ideal testing grounds for further work in this area.

\section{Acknowledgments}

This work was supported in part by the Natural Sciences and Engineering
Research Council. We would like to thank K. Lake for helpful discussions.

\end{document}